\documentclass[12pt]{article}%
\usepackage{amsfonts}
\usepackage{bm}
\usepackage{cite}
\usepackage{mathrsfs}
\usepackage{amsmath}
\usepackage{graphicx}
\usepackage{hyperref}
\usepackage{verbatim}
\usepackage{color}
\usepackage{enumerate}
\usepackage{amsfonts}
\usepackage{amssymb}
\usepackage{ulem}
\usepackage{authblk}
\usepackage[hang,flushmargin]{footmisc}
\usepackage{lipsum}
\usepackage[font=footnotesize]{caption}
\usepackage{soul}
\usepackage[utf8]{inputenc} 
\usepackage{empheq}

\csname @addtoreset\endcsname{equation}{section}
\newcommand{\eal}[1]{\begin{equation} \begin{aligned} #1 \end{aligned}\end{equation}}

\newcommand{\badat}{\begin{alignedat}} 
\newcommand{\eadat}{\end{alignedat}}

\usepackage[T1]{fontenc}
\usepackage{tikz}
\usetikzlibrary{arrows,calc,positioning}

\tikzstyle{intt}=[draw,text centered,minimum size=6em,text width=5.25cm,text height=0.34cm]
\tikzstyle{intl}=[draw,text centered,minimum size=2em,text width=2.75cm,text height=0.34cm]
\tikzstyle{int}=[draw,minimum size=2.5em,text centered,text width=3.5cm]
\tikzstyle{intg}=[draw,minimum size=3em,text centered,text width=6.cm]
\tikzstyle{sum}=[draw,shape=circle,inner sep=2pt,text centered,node distance=3.5cm]
\tikzstyle{summ}=[drawshape=circle,inner sep=4pt,text centered,node distance=3.cm]

\begin{document}

\title{\vspace{-30pt} \Huge{\sc Celestial open strings at one-loop}\vspace{10pt}}

\author[a]{\normalsize{Laura Donnay}}
\author[b]{\normalsize{Gaston Giribet}}
\author[c]{\normalsize{Hern\'an Gonz\'alez}}
\author[d]{\normalsize{Andrea Puhm}}
\author[e]{\normalsize{Francisco Rojas}}

\affil[a]{{\small\textit{International School for Advanced Studies (SISSA),
Via Bonomea 265, 34136 Trieste, Italy}}}
\affil[ ]{{\small\textit{ Istituto Nazionale di Fisica Nucleare (INFN), Sezione di Trieste,
Via Valerio 2, 34127, Italy
}}}
\affil[b]{{\small\textit{Department of Physics, New York University. 726 Broadway, New York,
NY10003, USA}}}
\affil[c]{{\small\textit{Facultad de Artes Liberales, Universidad Adolfo Ib\'a\~nez, Santiago, Chile}}}
\affil[d]{{\small\textit{CPHT, CNRS, Ecole Polytechnique, IP Paris, F-91128 Palaiseau, France}}}
\affil[e]{{\small\textit{Facultad de Ingenier\'ia y Ciencias, Universidad Adolfo Ib\'a\~nez, Santiago, Chile}}}

\begin{flushright}
\texttt{CPHT-RR036.072023}
\end{flushright}

\begingroup
\let\newpage\relax
\date{}
\maketitle
\endgroup


\thispagestyle{empty}

\begin{abstract}
We study celestial amplitudes in string theory at one-loop. Celestial amplitudes describe scattering processes of boost eigenstates and relate to amplitudes in the more standard basis of momentum eigenstates through a Mellin transform. They are thus sensitive to both the ultraviolet and the infrared, which raises the question of how to appropriately take the field theory limit of string amplitudes in the celestial basis. We address this problem in the context of four-dimensional genus-one scattering processes of gluons in open string theory which reach the two-dimensional celestial sphere at null infinity. We show that the Mellin transform commutes with the adequate limit in the worldsheet moduli space and reproduces the celestial one-loop field theory amplitude expressed in the worldline formalism. The dependence on $\alpha'$ continues to be a simple overall factor in one-loop celestial amplitudes albeit with a power that is shifted by three units with respect
to tree-level, thus making manifest that the dimensionless parameter $g_{10}^2/\alpha'^3$ organizes the loop expansion in the celestial basis.

\end{abstract}

\newpage

\begin{small}
{\addtolength{\parskip}{-2pt}
 \tableofcontents}
\end{small}
\thispagestyle{empty}
\newpage

\section{Introduction}





A variety of old and new insights into low-energy aspects of gravity and gauge theories~\cite{Strominger:2017zoo} have reinvigorated the pursuit of a holographic formulation of quantum gravity in asymptotically flat spacetimes. Since the bulk Lorentz group acts as the global conformal group on the celestial sphere, the natural observables in flat space holography are celestial amplitudes which recast the standard S-matrix in a basis of boost (rather than energy) eigenstates~\cite{deBoer:2003vf,Cheung:2016iub,Pasterski:2016qvg,Pasterski:2017kqt,Pasterski:2017ylz}. They transform as correlation functions in an exotic conformal field theory (CFT) that lives on the celestial sphere at null infinity. This celestial CFT is conjectured to be holographically dual to quantum gravity in asymptotically flat spacetimes.

The approach towards establishing this celestial holography proposal so far has been predominantly bottom-up by exploring the implications of infrared aspects of gravity and gauge theories and by examining the structure of celestial amplitudes. 
This includes the identification of the symmetries of celestial CFT and their constraints on bulk scattering amplitudes~\cite{He:2014laa,Kapec:2016jld,Nande:2017dba,Donnay:2018neh,Fan:2019emx,Nandan:2019jas,Pate:2019mfs,Adamo:2019ipt,Puhm:2019zbl,Guevara:2019ypd,Kapec:2017gsg,Donnay:2020guq,Kapec:2021eug,Pasterski:2021fjn,Donnay:2022sdg,Pano:2023slc} and how basic CFT structures such as the operator product expansion (OPE) arise~\cite{Fan:2019emx,Pate:2019lpp,Ren:2022sws}.
This aspect of flat space holography does not require the detailed knowledge of a microscopic description such as string theory. Yet key new insights are to be expected from a string theory realization of celestial holography which remains an important outstanding goal. This puts the current status of flat space holography in contrast with the (much older and much better understood) AdS/CFT correspondence for which we have a string theory embedding~\cite{Maldacena:1997re,Aharony:1999ti} and this has led to a variety of crucial insights into the nature of quantum gravity and the formulation of a detailed holographic dictionary. 

The realization of such a top-down construction for asymptotically flat spacetimes -- see~\cite{Costello:2022wso,Costello:2022jpg,Costello:2023hmi} for interesting recent proposals in the context of twisted holography --   would allow us to ask new types of questions. One may, for example, wonder whether there exists a celestial counterpart of the $1/N$ expansion of AdS/CFT. This question is closely tied to the fundamental structure of celestial CFT since, e.g. non-analytic correlation functions may appear in OPEs as an artifact of a perturbative truncation. A step towards advancing this goal is to achieve a detailed understanding of celestial amplitudes in string theory.

A particularly appealing aspect of celestial amplitudes in string theory is their soft UV behavior which imposes interesting constraints on their analytic structure in the boost weight basis.
Celestial amplitudes in field theory, as defined by the Mellin transform of momentum space amplitudes, are in general not well-defined in the UV. However, as explored at tree-level in~\cite{Stieberger:2018edy}, the UV-softness of string theory\cite{Gross:1987ar,Gross:1989ge} renders the diverging energy integrals finite. This soft UV behavior of quantum gravity implies stringent constraints on the analytic structure of celestial amplitudes such that the exact four-particle scattering amplitude is meromorphic in the complex boost weight plane with poles confined to one side of the real axis\cite{Arkani-Hamed:2020gyp}. 
The expected exponential decay of string amplitudes at fixed-angle scattering arises at large real frequencies, but can have poles and branch cuts in the complex frequency plane 
whose imprint on celestial amplitudes
was studied in~\cite{Chang:2021wvv}.
Other interesting works on celestial string amplitudes include the computation of celestial OPEs from the string worldsheet in~\cite{Jiang:2021csc} and in the context of ambitwistor strings in~\cite{Adamo:2019ipt,Casali:2020uvr, Adamo:2021zpw,Bu:2021avc}. 
Many open questions remain and the study of celestial amplitudes in string theory deserves  more attention.

The first explicit results on celestial string amplitudes at tree-level are from the beautiful work~\cite{Stieberger:2018edy} of Stieberger and Taylor.
Their analysis of celestial 4-gluon and 4-graviton amplitudes in type I open superstring and closed heterotic superstring theories at tree-level revealed several interesting features. First, the UV-softness of string amplitudes renders potentially divergent Mellin integrals of field theory amplitudes finite and the resulting celestial amplitudes are well-defined. Second, the dependence on $\alpha'$ consists in a simple overall factor with a power that corresponds to the net conformal boost weight of the external particles. Third, they obtained the low-energy field theory limit of celestial string amplitudes at tree-level as a certain limit of the conformal cross ratio. Fourth, the analogue of the high-energy "super-Planckian" limit for celestial amplitudes was argued to correspond to the limit of large net boost weight and appears to pin the worldsheet of the string to the two-dimensional celestial sphere.

In light of the UV-IR mixing of celestial amplitudes these are particularly fascinating results. They also pose several new questions:
Does the simple dependence of celestial string amplitudes on $\alpha'$ persist at loop-level? If so, do the the overall powers of $\alpha'$ and the (dimensionful) ten-dimensional Yang-Mills coupling constant combine into a natural (dimensionless) genus expansion parameter?
Moreover, does the field theory limit of string amplitudes commute with the Mellin transform?
In this work we will answer these questions in the affirmative.

We will focus on 4-gluon scattering amplitudes in type I open string theory at 1-loop and consider configurations such that the external momenta are essentially four-dimensional while loop momenta are ten-dimensional. 
The celestial 4-gluon amplitude in Yang-Mills theory at tree-level  takes the form of an exotic two-dimensional correlation function which depends on the net boost weights of the external particles and has distributional support that enforces a reality condition on the conformal cross ratios. In string theory the celestial amplitude at tree-level includes the Mellin transform of the Veneziano amplitude. Such tree-level celestial string amplitudes were studied in~\cite{Stieberger:2018edy}.

We generalize their result beyond tree-level focusing on the planar (orientable) 1-loop contribution to 4-gluon scattering. Compared to tree-level the integral over the moduli is much more involved and so is the dependence on $\alpha'$. It is thus highly non-trivial that $\alpha'$-dependence at 1-loop continues to show up as a simple overall factor. Moreover, compared to tree-level the power of $\alpha'$ given by the net boost weight of the external gluons is shifted by three units - which is the precise power of $\alpha'$ needed to combine the (dimensionful) ten-dimensional Yang-Mills coupling constant $g_{10}$ into a dimensionless parameter $g_{10}^2/\alpha'^3$. It appears natural to identify this as the loop expansion parameter organizing different genus contributions of celestial amplitudes.

Another puzzling question for celestial string amplitudes concerns how to appropriately take the field theory limit and whether it commutes with the Mellin transform that mixes the high and low energy regimes. 
At loop-level the natural field theory limit arises from a limit on the moduli space regardless of the value of the conformal cross ratios of celestial amplitudes.
We compute this field theory limit for celestial 4-gluon amplitudes in string theory at 1-loop following two routes. First, we review how the field theory limit of string amplitudes of four gluons, corresponding to the appropriate sector of the moduli space of the genus-1 integrated correlators, matches the expression for the 4-gluon amplitude in Yang-Mills theory in the worldline formalism. Second, we perform the field theory limit directly on the celestial 1-loop string amplitudes and compare it to the Mellin transform of the Yang-Mills amplitude in the worldline representation. This demonstrates that the Mellin transform and the appropriate field theory limit in the moduli space of the string worldsheet CFT commute.

The paper is organized as follows. In section~\ref{sec:Preliminaries} we collect relevant formulae and notation and introduce celestial amplitudes. We review in section~\ref{sec:CelestialStringAmplitudes} the tree-level amplitude for four gluons in type I open superstring theory in the momentum and boost weight bases and discuss their forward scattering limit. In section~\ref{sec:CelestialStringLoop} we compute the 1-loop celestial string amplitude and show that, as for the tree-level amplitude, the dependence on $\alpha'$ is an overall factor albeit with a shifted power. In section~\ref{sec:Worldline} we discuss the field theory limit of celestial string amplitudes at one-loop.
We conclude in section~\ref{sec:Conclusions} with some open questions.

\section{Preliminaries}
\label{sec:Preliminaries}

A generic 4-point scattering amplitude in momentum-space is a function of the Mandelstam invariants $s, t, u$ defined in terms of the external particle momenta $p_j$ as\footnote{We use the mostly-plus metric and the kinematic conventions of \cite{Green:2012oqa,Green:2012pqa}.}
\begin{equation}
\begin{split}
s = -(p_1 + p_2)^2, \quad
t = -(p_2 + p_3)^2, \quad
u = -(p_1 + p_3)^2.
\end{split}
\end{equation}
Momentum conservation and $p_i^2=0$ implies $s + t + u = 0$ so that any massless 4-point scattering amplitude $\mathcal{A}_4$ can be written as 
\begin{equation}
\label{eq:mssa}
\mathcal{A}_4(p_1, p_2, p_3, p_4) = A(s,t,\{J_i\}) \, \, \delta^{(D)}(p_1 + p_2 + p_3 + p_4 )\,,
\end{equation}
where $J_i$ are the helicities of each of the external particles.
We can parametrize the momenta for each external massless state $p_j^\mu$ as
\begin{equation}
\label{celestialp}
p_j^\mu = \eta_j \, \omega_j\, q^\mu(\vec{z})\,,
\end{equation}
where $\omega_j$ is the (positive) frequency of the $j^{\text{th}}$ particle, $\vec{z}$ labels a point on the $d=D-2$ dimensional celestial sphere, $\eta_j=\pm 1$ denotes incoming versus outgoing particles, and $q^{\mu }$ is a null vector pointing towards the celestial sphere
\begin{equation}
q^\mu(\vec{z}_j) = (1+|\vec{z}_j|^2, 2\vec{z}_j,1-|\vec{z}_j|^2)\, .
\end{equation}
Here, $i,j=1,2,3,4$, $\mu=0,1,2,\dots,D-1$; see the appendix for conventions. 

In this work we are interested in 4-gluon amplitudes in open superstring theory, their field theory limits and their representations as celestial amplitudes on the 2-sphere at null infinity. To that end we will consider momentum configurations $p^\mu_j=(p^0_j,p^1_j,p^2_j,p^3_j,\vec{0})$, while loop momenta $\ell^\mu$ are ten-dimensional. 

The celestial amplitude, $\widetilde{\mathcal A}_4(\{\Delta_i,\vec{z}_i\})$, is obtained from the momentum space amplitude, $\mathcal{A}_4(\{p_i\})$, by performing a Mellin transform with respect to the external frequencies, namely
\eal{
\label{eq:Mellin}
\widetilde{\mathcal A}_{4}(\{\Delta_i,\vec{z}_i\}) &=   \,\int_0^{\infty  } \prod_{i=1}^4\Big( \omega_i^{\Delta_i-1} d\omega_i \Big)\, \, \delta^{(4)}( p_1^\mu+p_2^\mu+p_3^\mu+p_4^\mu ) \, \,A(\{\omega_i, \vec{z}_i\})\,.
}
These celestial amplitudes transform as $2$-dimensional correlation functions with conformal boost weights $\Delta_i=1+i\lambda_i$ where $\lambda_i\in \mathbb{R}$.
It will be convenient to define the net boost weight $\beta$ and denote the scattering angle $\theta$ in the center-of-mass frame through 
\begin{equation}
\label{r-angle}
    \beta=-\frac{i}{2}\sum_{j=1}^4 \lambda_j, \quad \quad r=-\frac{s}{t}=\csc^2\left(\frac{\theta}{2}\right)\,.
\end{equation} 
Using the parametrization \eqref{celestialp}, we can express $\vec z_j=(z_j,\bar z_j)$ in complex coordinates and, after enforcing momentum conservation, write
\begin{equation}
r= \frac{z_{12}z_{34}}{z_{23}z_{41}} \, ,  \ \ \ \ \ \ \bar r= \frac{\bar z_{12}\bar z_{34}}{\bar z_{23}\bar z_{41}}\,.\label{Elr} 
\end{equation}
where $z_{ij}\equiv z_i-z_j$. These represent cross ratios in the $(z,\bar z )$ plane.

Momentum conservation in the 4-point case allows for three of the four Mellin integrals in \eqref{eq:Mellin} to be easily evaluated and implies that the resulting celestial amplitude is distributional with support on $r = \bar{r}$. In order to  explicitly evaluate the Mellin transform, it is convenient to factorize the amplitude as $A(s,t,\{J_i\})=R(\{ z_i, \bar{z}_i,J_i\})B(s,t)$, where $B$ only depends on Mandelstam variables and $R$ is a rational function that carries the helicities of the external particles, but whose precise expression is not relevant for the discussion presented here. The celestial 4-point amplitude then takes the form \cite{Gonzalez:2020tpi}
\begin{equation}
\label{eq:gca}
\widetilde{\mathcal{A}}_4(\{\Delta_i, z_i, \bar{z}_i\}) =2 K(\{\Delta_i, z_i, \bar{z}_i\})\, r^{\frac{5-\beta}{3}}(r-1)^{\frac{2-\beta}{3}} \, \delta(r - \bar{r}) \Theta(r-1){\tilde{B}}(r,\beta)\,, 
\end{equation}
with ${\tilde B}$ being the Mellin transform of the form factor, namely
\begin{equation}
\label{eq:gca2}
{\tilde{B}}(r,\beta)=\int_0^{\infty} d\omega \,\omega^{-\beta - 1} B( r \omega, -\omega)\,,
\end{equation}
where 
\eal{
 K(\{\Delta_i, z_i, \bar{z}_i\}) = \prod_{i=1}^4 z_{ij}^{h/3-h_i-h_j} {\bar z}_{ij}^{\bar{h}/3-\bar{h}_i-\bar{h}_j}
}
has the transformation properties of a 4-point correlation function in a $d=2$ CFT.
Here 
\eal{
h_i= \tfrac{1}{2}\left(\Delta_i+J_i\right)  \quad \bar{h}_i= \tfrac{1}{2}\left(\Delta_i-J_i\right)\,,
}
and $h=\sum_{i=1}^4 h_i$, $\bar h=\sum_{i=1}^4 \bar h_i$. 

Note that the integral in $\tilde{B}$ only depends on the conformally invariant cross-ratio $r$ and the sum of all scaling dimensions though $\beta$. The step function $\Theta(r-1)$ simply enforces the condition for physical scattering $r>1$. {Nevertheless, one can also consider the analytic continuation of the expression to the $r<0$ region of the kinematic space; for example, this is considered in \cite{Stieberger:2018edy} and it will also be considered in the integrals appearing throughout this article. }

\section{Celestial string amplitudes}
\label{sec:CelestialStringAmplitudes}

In this section we review celestial string amplitudes at tree-level and their forward scattering limit as studied in~\cite{Stieberger:2018edy}. We end with some remarks about this limit which will be relevant for the study of celestial string amplitudes at 1-loop.

\subsection{Review of string amplitudes in momentum space}

We will be concerned with superstring amplitudes up to 1-loop order, i.e.
\begin{equation}
    \mathcal A_{\rm string}=\mathcal A^{(0)}_{\rm string}+\mathcal  A^{(1)}_{\rm string}+\dots
\end{equation}
where $(0)$ indicates the tree-level contribution and $(1)$ the 1-loop correction. The (color-ordered) 4-gluon  amplitude in type I string theory, including the tree-level and its 1-loop correction, is given by
\begin{equation}
\label{eq:tree_loop}
     A_{\rm string}(p_1,p_2,p_3,p_4)={A}_{YM}^{(0)}(\{p_i,J_i\}) \left(\, f^{(0)}\,  +\,  f^{(1)}\, +\, \dots \right),
\end{equation}
with $f^{(0)}, \, f^{(1)}, ...$ being contributions from tree level and different loops in the expansion. $f^{(0)}$ is the Veneziano amplitude
\begin{equation}
f^{(0)}(s,t) = \frac{\Gamma(1-\alpha's)\Gamma(1-\alpha't)}{\Gamma(1-\alpha's-\alpha't)}\, ,\label{arribeno2}
\end{equation}
which for ${\rm Re}(\alpha' t)<1$, ${\rm Re}(\alpha' s)<0$ and using the integral representation of the Euler beta function can be expressed as
\begin{equation}
     f^{(0)}(s,t) = -\alpha' s \int_0^1dx\, x^{-\alpha' s-1}(1-x)^{-\alpha' t}\, .\label{eq:EulerBeta1}
\end{equation}
This representation naturally arises from the exponentiation of propagators on the disk when Wick contracting the vertex operators.

The 1-loop contribution $f^{(1)}$ has a planar, orientable piece given by\footnote{For a very recent development on the explicit evaluation of this expression, see \cite{Eberhardt:2023xck}.}
\begin{equation}
     f_P^{(1)}(s,t)=\frac{16 \pi^3 g_{10}^2 \, st}{\alpha'} \int_0^1 \frac{dq}{q} (F_1(a,q))^{10-D}  \int_0^1 \prod_{r=1}^3  d\nu_r \Theta (\nu_{r+1}-\nu_r) \left(\frac{\psi_{12}\psi_{34}}{\psi_{13}\psi_{24}}\right)^{-\alpha's}\left(\frac{\psi_{23}\psi_{14}}{\psi_{13}\psi_{24}}\right)^{-\alpha't}\label{arribeno}\,.
\end{equation}
Here, $g_{10}$ is the 10-dimensional Yang-Mills coupling constant and $F_1$ is the factor that arises in toroidal compactification to $D<10$ with compactification radii $R=(\alpha')^{1/2}/a$. The dependence on the functions $\psi_{ij}=\psi(\pi(\nu_{i}-\nu_{j}),q)$ -- see \eqref{ThePsi} below for its explicit expression in terms of Jacobi functions -- follows from 
\begin{equation}
\badat{2}
 \prod_{1\leq i< j\leq 4} (\psi_{ij})^{2\alpha' p_i\cdot p_j}=\left(\frac{\psi_{12}\psi_{34}}{\psi_{13}\psi_{24}}\right)^{-\alpha's} \left(\frac{\psi_{14}\psi_{23}}{\psi_{13}\psi_{24}}\right)^{-\alpha't}\,,
\eadat
\end{equation}
where we are using the definition of the Mandelstam variables, the massless condition for the external states, and momentum conservation $s+t+u=0$. This comes from the exponentiation of propagators on the annulus when Wick contracting the vertices. 

The ellipsis in \eqref{eq:tree_loop} denote further quantum corrections including those coming from 1-loop non-orientable diagram i.e. the Moebius strip diagram, which ensures UV finiteness of open string amplitudes at one loop, as well as from non-planar insertion configurations. We will comment on these contributions in the next subsection.

Notice from \eqref{eq:tree_loop} that both string tree-level and 1-loop amplitudes are proportional to the Yang-Mills field theory amplitudes at tree-level. In particular this is true even for the non-planar contribution, and it is believed to hold at higher genus as well. Thus, all the information of the particles' helicities is encoded in the tree-level field theory amplitude, while all the stringy dependence lies in the expressions inside the parentheses in \eqref{eq:tree_loop} which only depend on the Mandelstam invariants $s_{ij}=-(p_i+p_j)^2$.

\subsection{Celestial string amplitudes at tree-level}\label{ssec:TreeCelestialString}

At tree-level, the MHV amplitude is the only non-vanishing one; hence, we focus on that type of amplitudes here. From the factorization in \eqref{eq:tree_loop} we see that the tree-level 4-gluon MHV amplitude in type I string theory is 
\eal{
\label{4gluon-string}
A_{\rm string}^{(0)} (-,-,+,+)= A_{YM}^{(0)}(-,-,+,+) f^{(0)}(s,t)
}
with the (color-ordered) 4-gluon Yang-Mills amplitude given by
\eal{
\label{eq:4MHV}
A_{YM}^{(0)}(-,-,+,+) = g_{10}^2\,\frac{\omega_1 \omega_2}{\omega_3 \omega_4} \frac{z_{12}^3}{z_{23} z_{34} z_{41}} = g_{10}^2\, r\, \frac{z_{12}\bar z_{34}}{\bar z_{12} z_{34}}\,.
} 
The corresponding celestial amplitude obtained by a Mellin transform of each external particle is given by\cite{Stieberger:2018edy} 
\eal{
\label{Iintegral}
\tilde{f}^{(0)}(r,\beta) =  \int_0^\infty \omega^{-\beta-1}f^{(0)}(r\omega,-\omega) \, d\omega\,.
}
Using~\eqref{eq:EulerBeta1} with $s=r\omega$ and $t=-\omega$ we find 
\eal{
\label{eq:Itree}
\tilde{f}^{(0)}(r,\beta) &=- \alpha'^\beta r \,\Gamma(1-\beta)\int_0^1 \frac{dx}{x}\,\left[r\log x-\log(1-x)\right]^{\beta-1}\,.
}
Notice that in writing \eqref{eq:EulerBeta1} we have used an integral representation for the Euler beta function which converges only if Re$(s)<0$, thus forcing us away from the physical scattering region. In the rest of this article we will assume that an analytic continuation back to Re$(s)>0$ is possible and must be performed at the end.

Expression (\ref{eq:Itree}) concludes the computation of the tree-level celestial string amplitude. Notice that all Mellin transforms have already been explicitly performed and the final result for $\tilde{f}^{(0)}(r,\beta)$ is indeed a function of the invariant cross-ratio $r$ and the total scaling dimension $\sum_{i=1}^4 \Delta_i$ through the $\beta$-dependence. 

\subsection{The limit of forward scattering}
\label{ssec:ForwardScattering}

In \cite{Stieberger:2018edy}, Stieberger and Taylor have argued that the tree-level celestial 4-gluon amplitude is recovered in the $r \to \infty$ limit. Using \eqref{r-angle} one can see that, from the bulk point of view, this corresponds to the forward limit of the 4-dimensional scattering process. In order to prepare the ground for the comparison with the 1-loop calculation, let us review this point in detail. We first expand $\tilde{f}^{(0)}(r,\beta)$ in powers of $1/r$, obtaining 
\eal{
\tilde{f}^{(0)}(r,\beta) &= - \alpha'^\beta r^{\beta}\Gamma(1-\beta)\int_0^1 \frac{dx}{x}\,(\log x)^{\beta-1} +\mathcal{O}(r^{\beta-1})\,,
}
which upon the change variables $x=\exp\left(-e^{y}\right)$ becomes 
\eal{
\tilde{f}^{(0)}(r,\beta) &= (-1)^\beta \alpha'^\beta r^{\beta}\Gamma(1-\beta)\int_{-\infty}^{\infty} dy \,e^{\beta y}+\mathcal{O}(r^{\beta-1})\,.
}
Since $\beta$ is purely imaginary, namely $\beta=-\frac{i}{2}\sum_{i}^4 \lambda_i$, we have
\eal{
\tilde{f}^{(0)}(r,\beta) &= 4\pi  \delta\Big(\sum_{i}^4 \lambda_i\Big)+\mathcal{O}(r^{\beta-1})
\label{I-FTlimit}
}
where we have enforced the condition $\beta=0$ from the delta function above. 
This shows that the $r \to \infty$ limit of the celestial correlator corresponding to the tree-level 4-gluon string amplitude~\eqref{I-FTlimit} studied in \cite{Stieberger:2018edy} exactly reproduces the celestial field theory amplitude \cite{Pasterski:2017ylz}.

We conclude this section by pointing out that the same answer could have been obtained by taking the opposite limit in the cross-ratio, that is, in the $r\to 0$ limit. In order to see this, we could either integrate by parts the $x$-integral in \eqref{eq:Itree} or, equivalently, we could have written the Euler beta function \eqref{eq:EulerBeta1} as
\eal{
f^{(0)}(r\omega,-\omega) = \alpha'\omega \int_0^1 x^{-\alpha'r\omega}(1-x)^{\alpha'\omega-1} \, dx \,.
\label{eq:EulerBeta2}
}
After making the change $x \to 1-x$ in the modular variable $x$ we obtain  
\eal{
\tilde{f}^{(0)}(r,\beta) = \alpha'^\beta \Gamma(1-\beta) \int_0^1 \frac{dx}{x} \left[r\log(1-x)-\log x\right]^{\beta-1}
}
from which we see that the field theory result \eqref{I-FTlimit} is now recovered by taking the $r\to 0$ limit. This duality is simply the manifestation of the crossing-symmetric form of the original (color-stripped) string amplitude under the change $s \leftrightarrow t$. This means that the $r\to \infty $ limit is not strictly necessary to reproduce the field theory approximation; this can also be achieved by other limit of the modulus. Indeed, the field theory limit of string amplitudes in celestial basis requires further study. In the case of 1-loop amplitudes, we will see that due to the presence of the moduli space of the genus-one surfaces the limit can be achieved for finite $r$.

\section{Celestial string amplitudes at one-loop}
\label{sec:CelestialStringLoop}

We will now compute the celestial 4-gluon open string amplitude at 1-loop.

\subsection{One-loop open string amplitudes}\label{sec:one-loop-string} 

The (color-stripped) MHV amplitude with four external massless vector open string states in the type I theory is 
\eal{
\label{eq:loop-amp}
A_{\rm string}^{(1)}(-,-,+,+)= A^{(0)}_{YM} (-,-,+,+)\, f^{(1)}(s,t)
}
where $A^{(0)}_{YM} (-,-,+,+)$ is the same kinematical expression appearing at tree-level \eqref{eq:4MHV} but now with the 1-loop stringy form-factor \cite{Green:1984ed}\footnote{See \cite{Green:2012pqa} equation (9.1.11) for the full expression with arbitrary polarizations and equation (10.4.10).}
\eal{
\label{eq:loop-F}
f^{(1)}(s,t)=  \frac{16 \pi^3 g^2_{10}}{\alpha' }st\int_0^1 \frac{dq}{q} \left[G(q^2)-G(-q^2)\right]+f^{(1)}_{NP}(s,t)\,.}
Here
$f^{(1)}_{NP}$ is the orientable, non-planar contribution, while the planar (orientable) contribution which we will focus on is
\eal{
\label{eq:loop-FP}
f^{(1)}_P(s,t)=  \frac{16 \pi^3 g^2_{10}}{\alpha' }st\int_0^1 \frac{dq}{q} G(q^2)\,,}
with
\begin{equation}\label{eq:loop-F2}
G(q^2) = \int_{\mathcal D} \prod_{i=2}^4 d\theta_i \prod_{i<j}\psi(\theta_{ji},q)^{2\alpha' p_i \cdot p_j}
\end{equation}
with $\theta_{ji}\equiv \theta_j-\theta_i$ and 
\begin{equation}\label{ThePsi}
\psi(\theta , q)= \sin\theta \prod_{n=1}^\infty \frac{1-2q^{2n}\cos2\theta+q^{4n}}{(1-q^{2n})^2} \, .
\end{equation} 

For computational convenience, we have absorbed a factor of $\pi$ in the argument of the exponentiated propagators $\psi(\theta_{ji},q)$. Now, the integral on the domain $\mathcal{D}$ in (\ref{eq:loop-F2}) means integrating $\theta_{2,3,4}$ in the interval $(0, \pi)$ keeping $\theta_1=0$ fixed, so that $\theta_{j1}=\theta_j$ and the order $0<\theta_2<\theta_3<\theta_4<\pi$. {Together with a factor that can be written in terms of the Jacobi function $\vartheta_1$, each exponential of the propagator contributes with a factor $-2\pi^2\log^{-1}q$ to the measure of integration on the modulus.}

Only the planar annulus and the M\"obius diagrams, $G(q^2)$ and $-G(-q^2)$, respectively contribute with the color factor ${\rm Tr(\lambda_{a_1}\lambda_{a_2}\lambda_{a_3}\lambda_{a_4})}$, whereas the 4-point non-planar annulus diagram $f^{(1)}_{NP}$ contributes with a color factor $\rm {Tr(\lambda_{a_1}\lambda_{a_2})\rm Tr(\lambda_{a_3}\lambda_{a_4})}$. A remarkable feature of the non-planar contribution is that it also factorizes out the contribution $A^{(0)}_{YM} (-,-,+,+)$. This is crucial for the string 1-loop computation as it guarantees the correct kinematic dependence of the amplitude. In the non-planar contribution, one deals with two insertions at different boundaries of the annulus. This amounts to considering both propagators between vertices in the same boundary, which are written in terms of the functions given in \eqref{ThePsi}, together with propagators between vertices at different boundaries, which involve the modified function  
\begin{equation}
\psi_{NP} (\theta , q)=  \prod_{n=1}^\infty \frac{1-2q^{2n-1}\cos2\theta+q^{4n-2}}{(1-q^{2n})^2} \, .
\end{equation}
Analogous to the planar contribution, {apart from a factor that can be written in terms of Jacobi function $\vartheta_4$, each exponential of the propagator contributes with a factor $-2\pi^2\log^{-1}q$.} For concreteness, hereafter we focus on the planar, orientable contribution. In the limit we will be dealing with, the other pieces contribute in a similar way.  

\subsection{Celestial string amplitudes at one-loop}
\label{ssec:CelestialStringAmplitudes1Loop}

The Mellin transform of the amplitude \eqref{eq:loop-amp} (with the momentum-conserving $\delta$-function reinstated) is controlled by the integral
\eal{
\label{eq:loop-energy-integral}
\tilde{f}^{(1)}(r,\beta)= \int_0^\infty d\omega \, \omega^{-\beta-1}f^{(1)}(r\omega,-\omega)\,.
}
It turns out that the Mellin transform of the planar contribution 
\begin{multline}\label{eq:f1P}
f^{(1)}_P(r\omega,-\omega) =-\frac{16 \pi^3 g_{10}^2}{\alpha'} r \omega^2 \int_0^1 \frac{dq}{q} \times \\
\int_{\mathcal{D}} \prod_{i=2}^4 d\theta_i   \Big(\frac{\psi(\theta_{42} , q)\psi(\theta_{3} , q)}{\psi(\theta_{43} , q)\psi(\theta_{2} , q)}\Big)^{-\alpha' r \omega}   \Big(\frac{\psi(\theta_{3} , q)\psi(\theta_{42} , q)}{\psi(\theta_{32} , q)\psi(\theta_{4} , q)}\Big)^{\alpha'\omega}
\end{multline}
can be evaluated exactly. Defining
\eal{\label{VW}
V \equiv \log  \left(\frac{\psi(\theta_{42} , q)\psi(\theta_{3} , q)}{\psi(\theta_{43} , q)\psi(\theta_{2} , q)}\right) \ \  , \quad  W \equiv \log   \left(\frac{\psi(\theta_{42} , q)\psi(\theta_{3} , q)}{\psi(\theta_{4} , q)\psi(\theta_{32} , q)}\right)\,,
}
and changing the order of integration we obtain  
\eal{
\tilde{f}^{(1)}_P(r,\beta)= - 16  \pi^3 g_{10}^2(\alpha')^{\beta-3} \Gamma(2-\beta)\, r\,
\int_0^1 \frac{dq}{q}  \int_\mathcal{D} \prod_{i=2}^4 d\theta_i \left[r V - W\right]^{\beta-2} \,,\label{Istring}
}
where we used the standard integral representation of the $\Gamma$-function. From this we observe that, as for the tree-level amplitude \cite{Stieberger:2018edy}, the entire dependence on $\alpha'$ for the celestial 1-loop amplitudes merely consists in an overall factor of powers of $\alpha'$. This observation will be crucial in the discussion of the field theory limit. Notice that, while the tree-level celestial amplitude with~\eqref{eq:Itree} 
is proportional to $\alpha'^{\beta }g^2_{10}$, the 1-loop result with \eqref{Istring} 
is proportional to $\alpha'^{\beta -3} g^4_{10}$. Indeed, this is consistent with the fact that the 10-dimensional Yang-Mills coupling constant $g_{10}$ has length dimension 3, so that $g^2_{10}/\alpha'^3$ is dimensionless. It seems natural to identify $g^2_{10}/\alpha'^3$ as the parameter that organizes the loop expansion in the celestial basis.

The $\beta$-dependence for tree-level celestial string amplitude captured by $\tilde{f}^{(0)}$ in \eqref{eq:Itree} is given as a representation in terms of an integral over the string moduli for a 4-point amplitude at tree-level, that is, the disk with four vertex operators at its boundary: the only modulus being the single real variable $x$ which is integrated over. Similarly for the 1-loop computation, $\tilde{f}^{(1)}$ here is also given in terms of an integral over the entire moduli of the 4-point open string amplitude at 1-loop, i.e. the annulus and the M\"obius strip with four vertex operator insertions at their suitable boundaries.\footnote{The planar amplitude (the annulus), represented by the functions $V(q^2)$, has the vertex insertions at its outer boundary only, while the non-orientable one (the M\"obius strip), represented by the functions $V(-q^2)$, has only one boundary with the vertex operators inserted on it.}

It is amusing to note that the argument of the $\Gamma$-function and the exponent of the integrand in~\eqref{Istring} is shifted by one unit compared to~\eqref{eq:Itree}. It would be interesting to understand how the residues of the poles at tree-level get modified at 1-loop. This deserves further study.


\section{Field theory limit and the worldline formalism}
\label{sec:Worldline}

In this section we study the field theory limit of celestial 1-loop open string amplitudes using the worldline formalism. 

\subsection{Low energy limit in momentum space}
\noindent We begin with a brief review of the field theory ($\alpha' \to 0$) limit of the 1-loop string amplitude in momentum space. We will show that it naturally yields the 1-loop amplitude in field theory when written in the language of the worldline formalism.
Let us first rewrite
\eal{
\label{appexp}
\prod_{i<j} \psi(q,\theta_{ji})^{2\alpha' p_i \cdot p_j} = \exp\left\{\sum_{i<j} 2\alpha' p_i \cdot p_j \log \psi(q,\theta_{ji})\right\}.
}
It is worth mentioning that this factor appears in all string amplitudes at 1-loop due to the exponential dependence on the Mandelstam invariants $p_i\cdot p_j$ in the vertex operators for all possible external states. 

The field theory limit is controlled by the region $q\to 1$ (from below). In order to study this regime, it is convenient to make the change of variables
\eal{
w = \exp\left\{\frac{2\pi^2}{\log q}\right\},
}
and study the $w\to 0$ limit instead. This is depicted in the Figure 1.
\begin{figure}[ht]
\begin{center}
\includegraphics[width=0.5\textwidth]{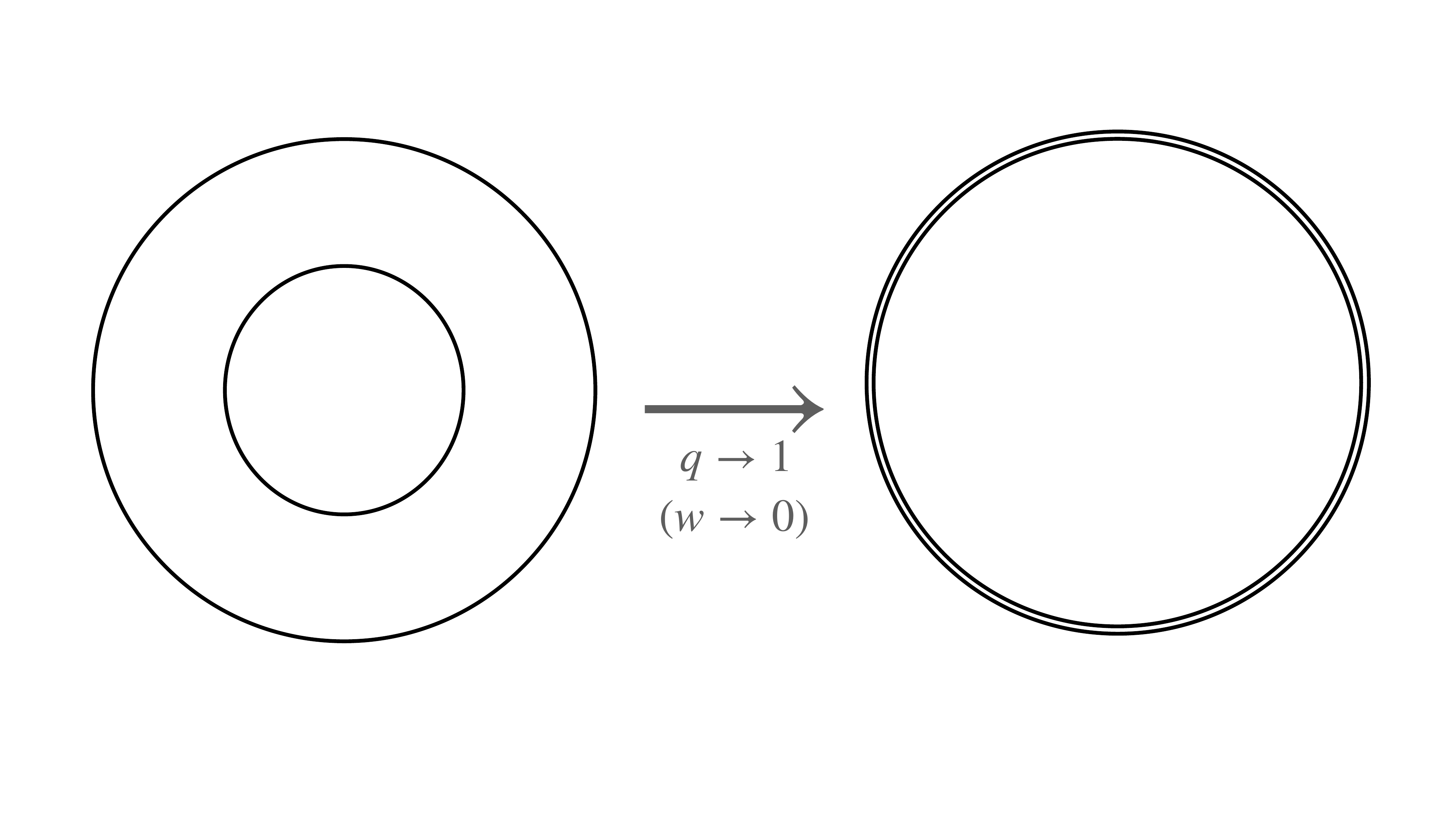}
\end{center}
\vspace{-35pt}
\caption{Point-particle approximation of the 1-loop string diagram.}
\end{figure}
Up to subleading order in the $w\sim 0$ expansion (including a purely $q$-dependent prefactor in $\psi$), one obtains \cite{Green:2012pqa} 
\eal{
\log \psi_{ji} = - \theta_{ji}(\pi -\theta_{ji}) \frac{\log w}{2\pi^2} +\log\left(\frac{\pi}{-\log w}\right)+ \mathcal{O}(w).
\label{smallwPsi}
}
Using this in the exponent of \eqref{appexp} yields
\eal{
\sum_{i<j} 2\alpha' p_i \cdot p_j \log \psi_{ji}= \sum_{i<j} 2\alpha' p_i \cdot p_j \left[ - \theta_{ji}(\pi -\theta_{ji}) \frac{\log w}{2\pi^2} + \log\left(\frac{\pi}{-\log w}\right)+\mathcal{O}(w)\right]
}
The leading term in the small $w$ expansion above vanishes due to momentum conservation and the massless condition $p_i^2=0$, since $\sum_{i<j} p_i \cdot p_j=0$; therefore
\eal{
\prod_{i<j} \psi_{ji}^{2\alpha' p_i \cdot p_j} = \exp\left\{-\frac{\alpha' }{\pi^2}\log w \sum_{i<j} p_i \cdot p_j \, \theta_{ji}(\pi-\theta_{ji})+\mathcal{O}(w)\right\}\,.
\label{smallw}
}
The field theory limit of the 1-loop string amplitude is thus
\begin{multline}
\label{Astring-thetasq-limit}
A^{(1)}_{\rm string} \underset{\, q\to 1}{\simeq } \frac{32\pi^5 g_{10}^2}{\alpha ' } A_{YM}^{(0)}\,st\int_0^{\varepsilon} \frac{dw}{w} \frac{1}{\log^2 w} \times \\
\int_{0}^{\pi }d\theta_4
\int_{0}^{\theta_4}d\theta_3
\int_{0}^{\theta_3 }d\theta_2  
\,\, \exp\left\{-\frac{\alpha' }{\pi^2}\log w \sum_{i<j} p_i \cdot p_j \, \theta_{ji}(\pi-\theta_{ji})
\right\}\,,
\end{multline}
which is of order $g_{10}^4/\alpha'$ since $A_{YM}^{(0)}\sim g_{10}^2$. In order to focus on the $w\sim 0$ contribution only, we have also introduced the cutoff $\varepsilon$ which is arbitrarily small but held away from zero.

Notice also that, since $w \ll 1$, the factor $\log w$ in the exponent blows up. However, the field theory limit is achieved by taking $\alpha' \to 0$ as in this limit all the massive string states become infinitely heavy leaving only the massless part of the spectrum. Thus, we need to take the joint limit $w \to 0$, $\alpha' \to 0$ but keeping the product $\alpha' \log w$ finite. After the change of variables 
\eal{
\label{T}
T  = -\frac{\alpha' \log w}{2\pi^2}\,,
}
we have
\begin{multline}
A_{\rm string}^{(1)} \underset{\, q\to 1}{\simeq } {16 \pi^3 g_{10}^2} \, A_{YM}^{(0)}\, st \int_{-\frac{\alpha'}{2\pi^2}\log \varepsilon}^{\infty} \frac{dT}{T^2} \times \\
\int_{0}^{\pi }d\theta_4
\int_{0}^{\theta_4}d\theta_3
\int_{0}^{\theta_3 }d\theta_2  
\,\,  \exp\left\{T \sum_{i,j=1}^4 p_i \cdot p_j \, \theta_{ji}(\pi-\theta_{ji})
\label{La57}
\right\}
\end{multline}
where $\theta_1=0$. Notice that the overall factor no longer depends on $\alpha '$. Now, we send $\alpha' \to 0$ while holding $\varepsilon$ fixed (we will remove this regulator at the end). This effectively sets the $T$-integration over the full range $(0,\infty)$ which is precisely what is needed in order to arrive at the field theory amplitude.

To precisely match with the field theory result, let us first make the following changes of variables
\eal{
\theta_4= \frac{\pi}{T} \tau_1 \,, \quad\theta_3= \frac{\pi}{T} \tau_2  \,,\quad \theta_2= \frac{\pi}{T} \tau_{3} \,.
}
In the case of four massless external states we get 
\begin{multline}
 \lim_{\substack{\alpha' \to 0}} A^{(1)}_{\rm string} \underset{\, q\to 1}{\simeq } 16{\pi^3}g_{10}^2 A_{YM}^{(0)}\, st \int_0^\infty \frac{dT}{T^5}
 \times \\ \int_0^T d\tau_1 \int_0^{\tau_1} d\tau_2 \int_0^{\tau_2} d\tau_3 \, \exp \left\{ \sum_{i,j=1}^4 \pi^2 G_{ij}\, p_i\cdot p_j\right\}
\label{Astring-thetasT-limit-2}
\end{multline}
where $G_{ij}\equiv G(\tau_i,\tau_j)$ is given by
\begin{equation}
\label{GG}
G(\tau_i,\tau_j) = |\tau_i-\tau_j| -\frac{(\tau_i-\tau_j)^2}{T}
\end{equation}
with $\tau_4=0$, and we have used momentum conservation. Equation \eqref{Astring-thetasT-limit-2} is precisely the expression describing the 4-point amplitude at 1-loop in $D=10$, expressed in the worldline formalism \cite{Schubert:2001he}, where $G_{ij}$ is, in fact, the worldline Green's function. Notice that, while we have done the computation in $D=10$, the same happens for arbitrary dimension $D$. To see that, one has to take into account that, in the $q\simeq 1$ limit, the compactification factor appearing in (\ref{arribeno}) has the asymptotic form $F_1(a,q)\simeq (a/\sqrt{\pi })(-\log w)^{1/2}+\cdots$, so that the integrand in the worldline formalism acquires an extra factor of the Schwinger parameter, namely $F_1(a,q)\sim T^{5-D/2}$, which is in agreement with the formulae in \cite{Schubert:2001he}. This manifestly shows that the Yang-Mills expression is correctly reproduced if the limit $\alpha'\to 0$, $q\to 1$ is taken properly. In the remainder of this section we will study how this limit of the 1-loop amplitudes works in the celestial basis.

\subsection{Field theory limit in celestial basis and worldline formalism}
\label{ssec:FieldTheoryCelestialWorldline}

The Mellin transform of \eqref{Astring-thetasT-limit-2} (again with the momentum-conserving $\delta$-function reinstated) yields the celestial 4-gluon amplitude in (super-)Yang-Mills theory at 1-loop. It will be convenient to define the functions $\tilde{G}_{ij}\equiv \frac{\pi^2}{T}G\left(\frac{T}{\pi} \theta_i,  \frac{T}{\pi} \theta_j \right)$, which are independent of $T$, in terms of which the exponent in \eqref{Astring-thetasT-limit-2} can be written as
\eal{
\label{GC}
\sum_{i,j=1}^4 \pi^2 G_{ij}\, p_i\cdot p_j\equiv -T C  t 
}
with
\eal{
\label{GC2}
C=r (\tilde{G}_{12}+ \tilde{G}_{34}-\tilde{G}_{13}-\tilde{G}_{24})+ (\tilde{G}_{13}+\tilde{G}_{24}-\tilde{G}_{14}-\tilde{G}_{23})\,.
}
From \eqref{Astring-thetasT-limit-2} we get
\eal{
f_{P}^{(1)}(s,t) \underset{\, q\to 1}{\simeq } {16 \pi^3 g_{10}^2} \, st \int_{-\frac{\alpha'}{2\pi^2}\log \varepsilon}^{\infty} \frac{dT}{T^2} 
\int_{0}^{\pi }d\theta_4
\int_{0}^{\theta_4}d\theta_3
\int_{0}^{\theta_3 }d\theta_2  
\,\, 
\exp\left\{-T C t\right\}\,,
}
whose Mellin tranformed counterpart is
\begin{eqnarray}
\tilde{f}_P^{(1)}(r,\beta) \underset{\, q\to 1}{\simeq } -16\pi^3 g_{10}^2 \Gamma(2-\beta) \, r \,\int_{-\frac{\alpha'}{2\pi^2} \log \varepsilon}^{\infty } \frac{{dT}}{T^{4-\beta}} 
\int_{0}^{\pi }d\theta_4
\int_{0}^{\theta_4}d\theta_3
\int_{0}^{\theta_3 }d\theta_2  
\,(- C)^{\beta-2}\,. \nonumber \\ \label{CF1}
\end{eqnarray}
We can now integrate in $T$, yielding
\begin{equation}
\int_{-\frac{\alpha '}{2\pi^2}\log\varepsilon}^{\infty} \frac{dT}{T^{4-\beta}}\,=\, \frac{(-\frac{\alpha'}{2\pi^2} \log \varepsilon)^{\beta-3}}{3-\beta}.\label{CF2} 
\end{equation}
This is the 1-loop form factor in the celestial basis in the limit $q\to 1$ of the moduli space, where the genus-1 surface becomes a circle. In the limit $\alpha'\to 0$, the integral \eqref{CF2} diverges as ${\rm Re}(\beta-3)<0$. This reproduces the expected field theory UV divergence.

The same result can be obtained by first computing the Mellin transform of the 1-loop amplitude and then taking the field theory limit as we will show in the remainder of this section. The fact that the Mellin transform and the field theory limit commute is a priori not obvious as celestial amplitudes are sensitive to both the ultraviolet and the infrared.

\subsection{Field theory limit of celestial string amplitudes} 
\label{FieldTheoryCelestialString}

We now perform directly the $q\to 1$ limit of the celestial string amplitude at 1-loop controlled by \eqref{Istring}
which we repeat here for convenience
\eal{\label{Istring2} 
\tilde{f}^{(1)}_P = - 16 \pi^3 r  g_{10}^2(\alpha')^{\beta-3} \Gamma(2-\beta) 
\int_0^{1} \frac{dq}{q} \int_{\mathcal D} \prod_{i=2}^4 d\theta_i \left[r V - W\right]^{\beta-2}\,. 
} 
It will again be useful to make the coordinate change $\log q=2\pi^2/\log w$ and expand the functions $V$ and $W$ defined in~\eqref{VW} around $q \simeq 1$ ($w\simeq 0$), making use of \eqref{smallwPsi}, \eqref{T} and \eqref{GC}, which yields
\eal{
r V - W \underset{\, q\to 1}{\simeq } \frac{\log w}{2\pi^2}\,  C\,.
}
Inserting this into \eqref{Istring2} 
we find the $q\to 1$ limit 
\eal{
\tilde{f}^{(1)}_P 
\underset{\,q\to 1}{\simeq }
- 32\pi^5 r  g_{10}^2(\alpha')^{\beta-3} \Gamma(2-\beta) 
\int_0^{\varepsilon} \frac{dw}{w} \frac{1}{\log^2 w} \int_{\mathcal D} \prod_{i=2}^4 d\theta_i \left[\frac{\log w}{2\pi^2} \,  C\right]^{\beta-2}
\label{Istring22}
}
with $\varepsilon \simeq \exp(2\pi^2/\log q)$. 
Since the dependence on $\alpha'$ factors out we can now easily take the field theory limit. First we express~\eqref{Istring22} in terms of the Schwinger parameter $T$ given in~\eqref{T}, namely 
\begin{equation}
\tilde{f}^{(1)}_P = - 16  \pi^3 g_{10}^2 r \,\Gamma(2-\beta) 
\int_{-\frac{\alpha'}{2\pi^2}\log\varepsilon }^{\infty} \frac{dT}{T^{4-\beta}}  
\int_{0}^{\pi }d\theta_4
\int_{0}^{\theta_4}d\theta_3
\int_{0}^{\theta_3 }d\theta_2
\left(-C\right)^{\beta-2}\,.
\end{equation}
This expression does not depend on $\alpha'$ other than in the limit of integration; therefore, we can take the limit $\varepsilon\to 0$, which exactly reproduces the field theory result (\ref{CF1})-(\ref{CF2}).

In other words, the Mellin transform of the 1-loop Yang-Mills amplitude expressed in the worldline formalism matches exactly the $q\to 1$ limit of the finite $\alpha'$ expression for the celestial 1-loop string amplitude. A remarkable aspect of this is that the limits agree for all values of cross-ratio $r$.

\section{Conclusions}
\label{sec:Conclusions}

In this paper we studied celestial amplitudes in string theory beyond tree-level. We considered genus-1 scattering processes of massless vector bosons in open string theory and examined different aspects of these observables in the celestial basis.

One of the remarkable features of celestial string amplitudes, which was observed at tree-level by Stieberger and Taylor in \cite{Stieberger:2018edy}, is their simple dependence on $\alpha '$. In the case of 4-gluon scattering at tree-level, the dependence of the celestial string amplitudes on the string tension turns out to be an overall factor, $(\alpha')^{\beta}$. In section~\ref{ssec:CelestialStringAmplitudes1Loop} 
we have shown that a similar phenomenon persists at 1-loop where the dependence on the string tension is again an overall factor albeit with a shifted power, $(\alpha')^{\beta-3}$. 
This simple behavior can be regarded as a manifestation of the mixing of energy regimes: 
As expressed in \cite{Stieberger:2018edy}, in celestial amplitudes all the string excitations participate at the same footing; the peculiar dependence of $\alpha'$ at tree-level and 1-loop manifestly shows that. Our result, moreover, suggests that the quantity $g^2_{10}/\alpha'^3$ is the natural dimensionless parameter that organizes the genus expansion in the celestial basis. Note, that when compactifying down to $D$ dimensions, the corresponding quantity would be $g^2_{D}\alpha'^{2-D/2}$.

The fact that celestial string amplitudes exhibit a factorized dependence on $\alpha'$ raises the question as to how the field theory limit works. It was observed in \cite{Stieberger:2018edy} that even after integrating over the energies of the external particles there exists a limit in which the tree-level string amplitudes of four massless vector bosons approach the Yang-Mills result: this is the large-$r$ limit, namely the limit of forward scattering $\theta \simeq 0$. 

This statement is as interesting as it is intriguing for several reasons: Firstly, this seems to imply that the field theory limit is associated to particular kinematic configurations, and while one can argue that the large-$r$ limit is precisely the limit in which the process is dominated by the exchange of massless particles, it is still puzzling that for some values of the Mandelstam variables such a limit is not accessible. Secondly, in the forward scattering limit other amplitudes -- e.g. those involving gravitons -- become singular, and this requires a better understanding. Thirdly, it is not clear what such a limit would correspond to in the case of higher-point amplitudes, where the number of cross ratios increases. Even for the case of four gluons at tree-level we have shown in section~\eqref{ssec:ForwardScattering} that the field theory limit is also achieved for other values of the cross ratio $r$, which is a manifestation of crossing symmetry. 

Motivated by all these questions, we decided to investigate the field theory limit of celestial string amplitudes at 1-loop. We have shown in section~\eqref{ssec:FieldTheoryCelestialWorldline} that, if one takes the appropriate limit in the moduli space of genus-one correlators, the celestial string amplitude correctly reproduces the expression that is obtained by Mellin transforming the 1-loop Yang-Mills amplitudes expressed in the worldline formalism. Remarkably, we have found that this matching holds for all values of $r$. In addition, we have shown in section~\eqref{FieldTheoryCelestialString} that, while the celestial amplitudes are sensitive to both the ultraviolet and the infrared physics, the Mellin transform and the field theory limit of 1-loop amplitudes actually commute. 

Our results present a step towards advancing our understanding of celestial string amplitudes and the challenging goal of embedding celestial holography in string theory. Many questions remain open. An immediate question is how the field theory limit works for higher-point processes in the celestial basis, even at tree-level. Another question, with potentially deep implications is: what is the precise connection between the two-dimensional worldsheet CFT and the celestial CFT that is conjectured to be the holographic dual of quantum gravity in four-dimensional asymptotically flat spacetimes? The existence of such a connection was suggested in~\cite{Stieberger:2018edy} in the context of four-point amplitudes, but it remains an open question how such a connection would work at higher points and for higher genus. Even more puzzling is the question of what would become of such a relation in spacetime dimensions larger than four. A related open problem is how the genus expansion of string theory processes in the bulk is organized from the point of view of the celestial CFT. We leave these questions for future work.

\paragraph{Acknowledgements}
\[\]
 L.D. is partially supported by INFN Iniziativa Specifica ST\&FI. The work of F.R. has been supported by FONDECYT grants 1221920 and 1211545. H.G. is partially supported by FONDECYT grants 1230853 and 1210635. H.G. and F.R. have also been supported by ANID/ACT 210100 Anillo grant. A.P. is supported by the European Research Council (ERC) under the European Union’s Horizon 2020 research and innovation programme (grant agreement No 852386).

\appendix{ 

\section{Kinematic conventions}
We follow the conventions of \cite{Green:2012oqa}. We use $\eta_{\mu\nu}= \text{diag}(-,+,\dots,+)$ and the Mandelstam invariants 
\begin{equation}
\begin{split}
s = -(p_1 + p_2)^2, \quad
t = -(p_2 + p_3)^2, \quad
u = -(p_1 + p_3)^2,
\end{split}
\end{equation}
thus, using momentum conservation, for massless states we have $s+t+u=0$.

Using the parametrization \eqref{celestialp} for our case of interest, where the external particles reach the two-dimensional celestial sphere, we have
$$
2p_i\cdot p_j = -\eta_i \eta_j \omega_i \omega_j |z_{ij}|^2\,.
$$
Total momentum conservation is written as $\sum_i p_i^{\mu}=0$, thus, we need to choose some $\eta_i$'s positive and some negative. Here we choose $\eta_1=\eta_2=-\eta_3=-\eta_4=1$. With these conventions we have that
\eal{
s &= +\omega_1 \omega_2 |z_{12}|^2 = +\omega_3 \omega_4 |z_{34}|^2\\
t &= -\omega_2 \omega_3 |z_{23}|^2 = -\omega_1 \omega_4 |z_{14}|^2\\
u &= -\omega_1 \omega_3 |z_{13}|^2 = -\omega_2 \omega_4 |z_{24}|^2\,.
}
Solving the $\sum_i p_i^{\mu}=0$ system of equations, yields
\eal{
\label{4system}
\omega_1&= \frac{z_{24}\bar z_{34}}{z_{12} \bar z_{13}} \omega_4 \,,\,\quad\, \omega_2&= \frac{z_{14}\bar z_{34}}{z_{12} \bar z_{32}}\omega_4 \,,\,\quad\, \omega_3&= \frac{z_{24}\bar z_{41}}{z_{23} \bar z_{13}}\omega_4 \,,\,\quad\, \frac{z_{12}z_{34}}{ z_{32} z_{14}}=\frac{\bar z_{12}\bar z_{34}}{\bar z_{32}\bar z_{14}}\,.
}
Using these, we can write the dimensionless parameter $r=-s/t$ in terms of the $z_i$ variables, i.e.
\eal{
r= - \frac{s}{t} = \frac{\bar z_{12}\bar z_{34}}{\bar z_{32}\bar z_{14}}\,.
}
Thus, from \eqref{4system}, we see that $r=\bar r$, which is consistent with the fact that, when written in the center-of-mass frame, one has indeed $r$ real, namely
\eal{
\frac{t}{s}= -\sin^2\left(\frac{\theta}{2}\right)
}
where $\theta$ is the scattering angle in this frame.

\bibliography{celestial-refs}
\bibliographystyle{utphys}

\end{document}